\documentclass[twocolumn,aps,prl,superscriptaddress]{revtex4-1}
\usepackage{amsmath,amssymb,amstext,graphicx,bm}

\usepackage{color}

\usepackage{cancel}

\begin{document}

\title{Zero field splitting of heavy-hole states in quantum dots}

\author{G. Katsaros}
\email{georgios.katsaros@ist.ac.at} \affiliation{Institute of Science and Technology Austria, Am Campus 1, 3400 Klosterneuburg, Austria}

\author{J. Kuku\v{c}ka}
\affiliation{Institute of Science and Technology Austria, Am Campus 1, 3400 Klosterneuburg, Austria}

\author{L. Vuku\v{s}i\'c}
\affiliation{Institute of Science and Technology Austria, Am Campus 1, 3400 Klosterneuburg, Austria}

\author{H. Watzinger}
\affiliation{Institute of Science and Technology Austria, Am Campus 1, 3400 Klosterneuburg, Austria}

\author{F. Gao}
\affiliation{National  Laboratory  for  Condensed  Matter  Physics  and  Institute  of  Physics, Chinese  Academy  of  Sciences,  100190,  Beijing,  China}

\author{T. Wang}
\affiliation{National  Laboratory  for  Condensed  Matter  Physics  and  Institute  of  Physics, Chinese  Academy  of  Sciences,  100190,  Beijing,  China}

\author{J. J. Zhang}
\affiliation{National  Laboratory  for  Condensed  Matter  Physics  and  Institute  of  Physics, Chinese  Academy  of  Sciences,  100190,  Beijing,  China}

\author{K. Held}
\affiliation{Institute of Solid State Physics, Vienna University of Technology, 1040 Vienna, Austria}

\date{\today{}}

\begin{abstract}
 Using inelastic cotunneling spectroscopy we  observe a zero field splitting within the spin triplet manifold of Ge hut wire quantum dots. The  states with spin $\pm1$ in the confinement direction
 are energetically favored  by   up to $55\mu$eV compared to  the spin 0 triplet state because of the  strong spin orbit coupling. The reported effect should be observable in a broad class of strongly confined hole quantum-dot systems and needs to be considered when operating hole spin qubits.
\end{abstract}

\maketitle

Hole states in semiconductor quantum dots  have gained increasing interest in the past few years as promising candidates for spin qubits due to their strong spin orbit coupling (SOC)~\cite{Winkler,Kloeffel2011,Hong2018}. Thanks to the SOC one now has a full-fledged electrical control of the hole spins~\cite{Maurand,Watzinger2018,Hendrickx2019,Crippa2019}, either via the electric-dipole spin resonance~\cite{Golovach}, g-tensor modulation~\cite{Kato}, or both \cite{Crippa2018}. Further, Rabi frequencies exceed 100MHz~\cite{Watzinger2018,Hendrickx2019} and reflectometry measurements reveal  spin relaxation times of $100\mu$s at $500mT$~\cite{Vukusic}, which underlies the big potential of hole spins as viable qubits. 

\begin{figure}
\includegraphics[width=1\columnwidth]{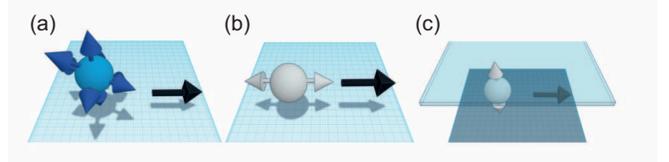}
\caption{
(color online) Spin physics for (a) electrons, (b) bulk holes, and (c)  confined HH states.  While for electrons the momentum (black arrows) and the spin are not correlated, for bulk holes the pseudospin is locked in the direction of motion because of their strong valence band SOC. By confining holes in two dimensions, HHs become energetically favorable and their pseudospin points in the confinement direction, perpendicular to their momentum. 
\label{Figure1}}%
\end{figure}

Despite the fact that a hole is simply a missing electron, their spins behave strikingly different than their electron counterparts~\cite{Kloeffel2018}. While the electron spin does not correlate with the direction of motion in typical semiconductors given their weak SOC [Fig.~\ref{Figure1}(a)]; the hole pseudospin points in the same direction as the momentum [Fig.~\ref{Figure1}(b)] already for bulk materials. This can be described by the 
Luttinger-Kohn Hamiltonian~\cite{LuttingerKohn, Luttinger} for holes near the $\Gamma$ point of the valence band, imposing a coupling between the momentum and the hole pseudospin.

 By introducing a strong confinement potential creating a quantum well, the heavy-hole (HH) light-hole (LH) degeneracy is lifted and the pseudospin changes its direction. For the HH states, which become energetically favorable, the pseudospin now points perpendicular to the momentum, i.e., in the direction of strong confinement [Fig.~\ref{Figure1}(c)] ~\cite{Kloeffel2018}. This implies that HHs confined in quasi two dimensional quantum dots (QDs), i.e., artificial atoms with strong confinement in one dimension, show spin anisotropy and could thus manifest similar effects as atoms show in the presence of a  magnetocrystalline anisotropy, i.e., a magnetic anisotropy leading to a zero field splitting (ZFS). However, to the best of our knowledge hitherto no  ZFS has been observed for quantum dots. 

For adatoms, on the other hand, ZFS studies have been at the focus of intense research as the magnetic anisotropy provides directionality and stability to the spin, which is the key for realizing nanoscale magnets. Scanning tunneling microscopy measurements have been used to reveal the magnetic anisotropy for several adatoms on surfaces and to understand how the local environment can influence it ~\cite{Rau,Jacobson,Miyamachi, Gambardella,Hirjibehedin}. ZFS as high as $58meV$, originating from the atomic spin orbit interaction, have been reported~\cite{Rau}.

Here, we use inelastic co-tunneling (CT)  to extract information about confined HH states. A hole-hole interaction strength of $275\mu$eV, similar to that of GasAs is reported. We have furthermore investigated the spin anisotropy of HH states confined in quasi two dimensional QDs. We measure a ZFS of up to $55\mu$eV for the excited triplet states confined in a QD with an even hole occupation. The evolution of the triplet states both for perpendicular and parallel magnetic fields is in very good agreement with the anisotropic Hamiltonian for the spin-triplet.

\begin{figure}
\includegraphics[width=8.6cm, keepaspectratio]{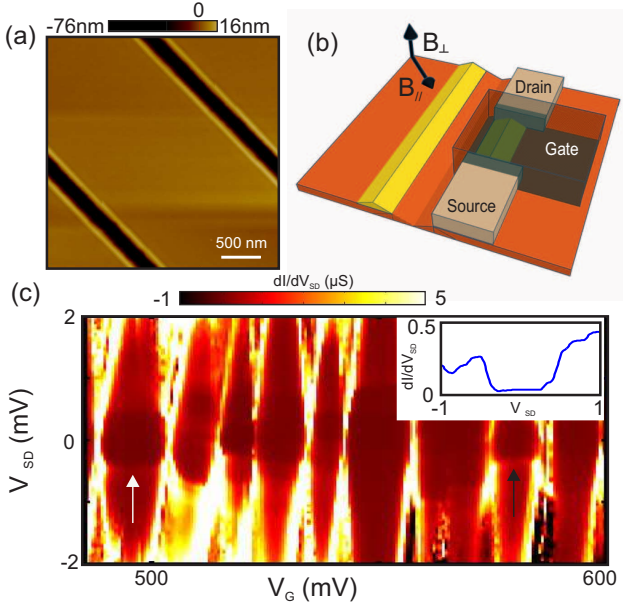}
\caption{\label{Figure2}
(color online) (a) Atomic force microscopy image showing parallel Ge HWs grown at the edges of trenches etched in the silicon wafers prior to growth. (b) Schematic showing the device geometry and the direction of the applied magnetic field. The HW is covered with a thin layer of hafnia (not shown), before the top gate is deposited. (c) Differential conductance $dI/dV_{SD}$ as a function of gate voltage $V_{G}$ and source-drain voltage $V_{SD}$ at $B= 0T$. The arrows indicate the position of two inelastic CT steps. The inset shows the $dI/dV_{SD}$ as a function of source-drain voltage $V_{SD}$ at the position of the white arrow.}
\end{figure}

The QDs used for this study are fabricated in Ge hut wires (HWs) grown by molecular beam epitaxy~\cite{Gao,Zhang}. These HWs are site-controlled as they are grown on Si wafers with predifined trenches [Fig.~\ref{Figure2}(a)]. The detailed description of the growth conditions can be found in Ref.~\cite{Gao}. They have a height of about 3.8nm and a width of approximately 38nm. Due to the strong confinement and compressive strain, the degeneracy between the HH and LH is lifted, leading to confined HH states~\cite{Katsaros2011,Watzinger2016}. The HWs are contacted by two $25$-nm-thick Pt electrodes, acting as source and drain, with a $50$-nm spacing between them. The gate electrode consists of $3$-nm Ti plus $25$-nm ${\rm Pt},$ and is separated from the source and drain contacts by hafnium oxide, deposited in  $80$ cycles of atomic layer deposition [Fig.~\ref{Figure2}(b)]. Two nominally identical devices from the fabricational point of view have been investigated in this study.

At low temperatures, transport through QDs is dominated by Coulomb blockade (CB), which leads to single electron transport. The stability diagram of a QD device with the characteristic Coulomb diamonds can be seen  in Fig.~\ref{Figure2}(c). However, due to second-order elastic CT processes the conductance within the coulomb diamonds does not drop to zero~\cite{Kogan,Franceschi}. At zero magnetic field, once the energy due to the bias voltage across the QD exceeds the orbital level separation, $\left|eV_{SD}\right|>E_{ORB}$,
the inelastic CT process leaves the QD in the excited orbital state
($e>0$ denotes the elementary charge).
The onset of inelastic CT is observed as a step in the differential conductance, $dI/dV_{SD}$, at $eV_{SD} = \pm E_{ORB}$ ~\cite{Franceschi,Kogan,Katsaros2010}, indicated by black arrows in Fig.~\ref{Figure2}(c).

\begin{figure}
\includegraphics[width=8.6cm, keepaspectratio]{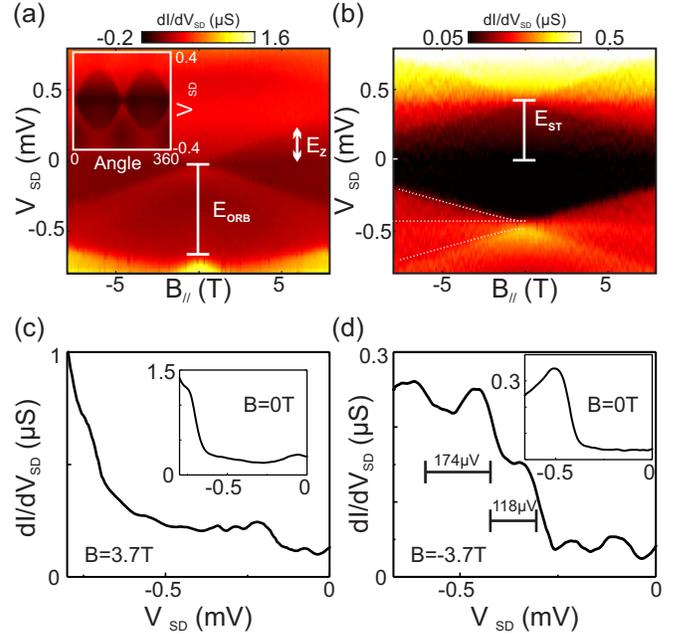}
\caption{\label{Figure3}
(color online) (a) $dI/dV_{SD}$ as a function of $B_{\parallel}$ and $V_{SD}$ at $V_{G}= 510.5mV$ for which the QD is in a spin-doublet ground state. Inset:  $dI/dV_{SD}$ at $B=1T$ as a function of $V_{SD}$ and magnetic field angle, from which a strong $g$-factor anisotropy  of about 7.5 ($g_{\parallel}=0.56 \pm 0.06$ and $g_{\perp}=4.17\pm 0.22$) can be extracted  similar to~\cite{Watzinger2016}. This anisotropy is due to the HH character of the confined states~\cite{Nenashev,Katsaros2011,Watzinger2016}. (b) $dI/dV_{SD}$ as a function of $B_{\parallel}$ and $V_{SD}$ at $V_{G}= 528.3mV$ for which the QD now has a singlet ground state with  $g_{\parallel}=0.57 \pm 0.01$ and $g_{\perp}=4.56\pm 0.16$. The dotted lines are the calculations from Fig.~\ref{Figure4}(b) (below) for $B_{\perp}\leq 0$, but reverted because the CT is at a negative $V_{SD}$ bias.(c,d) Line traces illustrating the CT steps at $B=3.7T$ and $B=-3.7T$, for the odd and even QD occupancy, respectively. The insets illustrate the $B=0T$ traces from which $E_{ORB}=690\mu\,$eV and $E_{ST}=415\mu\,$eV are extracted. The  unequal spacing between the triplet states in (d) of 174$\mu\,$eV - 118$\mu\,$eV reveals a ZFS of $55\mu$eV when rounded within our error of $5\mu$eV.}
\end{figure}

Inelastic CT is an excellent tool for magnetotransport spectroscopy measurements as the step width is not lifetime limited but depends only on the effective  temperature~\cite{Franceschi}. We first use it to extract information related to the strength of hole-hole interactions within a QD. When a QD confines an odd number of holes, the ground state is a (doubly degenerate) spin-doublet. On the other hand,  with an even number of holes the ground state of the  QD is a singlet state (assuming that the exchange coupling is weaker than the level splitting). Here, the two holes occupy the same (lowest in energy) orbital state with their pseudospins being antiparallel. The first excited states are the triplet states for which one hole occupies a higher orbital. 
This costs a higher energy for the  orbital occupation but also gains some 
 Coulomb repulsion energy compared to the singlet state  ~\cite{Hanson,Kouwenhoven}. By comparing the difference between the singlet-triplet energy $E_{ST}$ and the orbital level separation $E_{ORB}$, one can thus obtain  useful information about the strength of hole-hole interactions. 
 
\begin{figure*}
\includegraphics[width=17.2cm, keepaspectratio]{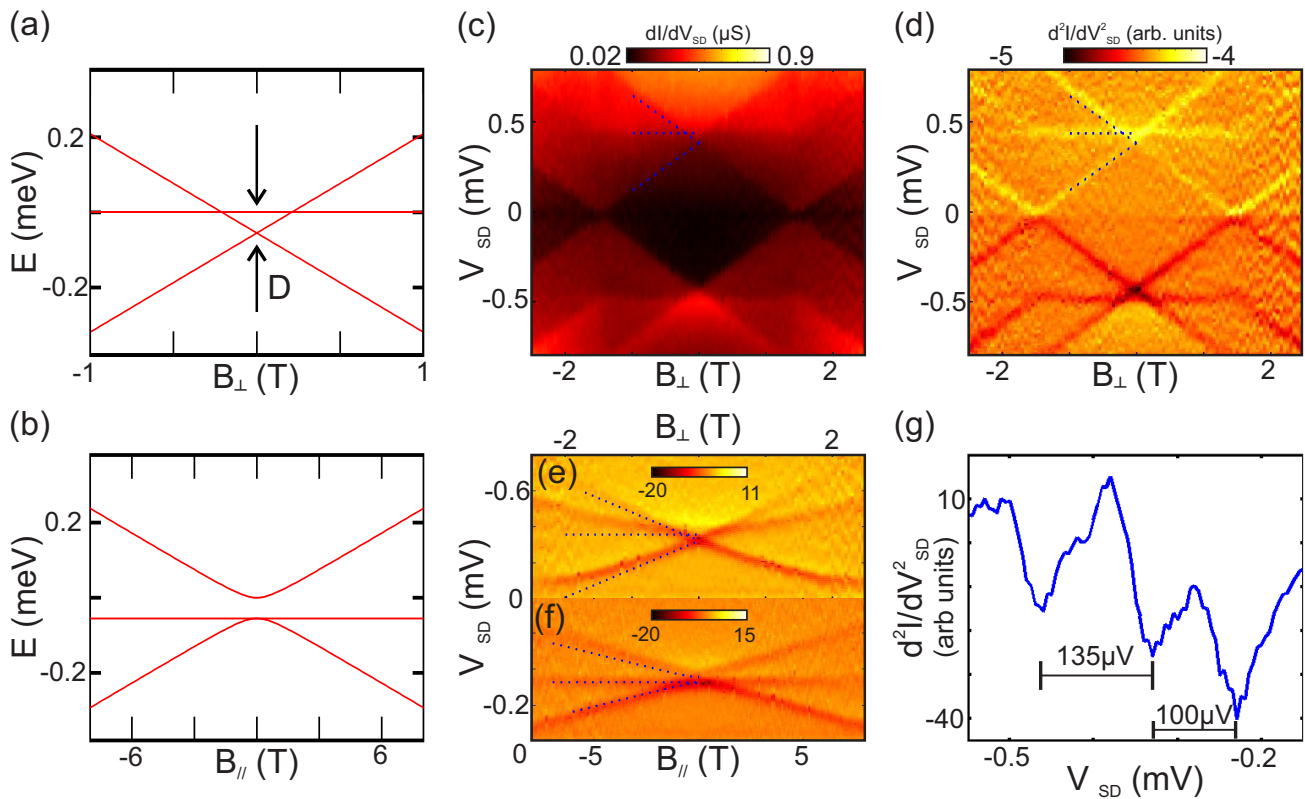}
\caption{\label{Figure4}
(color online) Evolution of HH triplet states from Hamiltonian (\ref{Eq}) for (a) $B_{\perp}$ and (b) $B_{\parallel}$,  using
$g_{\parallel}=0.57$, $g_{\perp}=4.56$, and $D=55\mu$eV, as extracted from the measurements when the QD is in the singlet ground state, see Fig.~\ref{Figure3}. (c) Experimental $dI/dV_{SD}$ and (d) numerical  derivative $d^2I/dV_{SD}^2$ as a function of $B_{\perp}$ and $V_{SD}$. The latter shows the ZFS more clearly; the calculation of (a) is plotted as dotted lines. (e,f) Numerical derivative $d^2I/dV_{SD}^2$  for a second device as a function of $V_{SD}$ and $B_{\perp}$, $B_{\parallel}$, respectively.  (g) Plot showing a line scan from (f) taken at $B_{\parallel}=-3.1T$ illustrating the unequal spacing of the triplet states. In this device a ZFS of $35\mu$eV $\pm$ $5\mu$eV is observed; again the dotted lines {in (e) and (f) are calculated based on the experimentally extracted parameters ($g_{\parallel}=0.52\pm 0.13$, $g_{\perp}=2.78\pm 0.06$, and $D=35\mu$eV) without any additional fitting. The discrepancy in (e) is due to orbital effects which are not taken into account in the model.}}
\end{figure*}

In order to conclude about the even/odd occupancy of the QD we investigate the evolution of the CT steps. For an odd number of holes, a magnetic field $B$ lifts the spin degeneracy of the doublet state by the Zeeman energy $E_{Z} = g \mu_B B$, where $g$ and $\mu_B$ are the hole g-factor and Bohr magneton, respectively. Once the energy due to the bias voltage across the QD exceeds the Zeeman energy, $\left|eV_{SD}\right|>E_{z}$,
the inelastic CT processes can flip the QD spin, leaving the QD in the excited spin state. This is visible as a step in Fig.~\ref{Figure3}(a). For zero magnetic field the observed feature corresponds to the first orbital excited state from which an orbital level separation $E_{ORB}$ of $690\mu$eV can be extracted. For an even number of holes, on the other hand, the magnetic field should split the three triplet states, and three inelastic cotunneling steps should be observed (note that the other state involved, the ground state singlet, does not split in a magnetic field). Indeed this behavior can be observed in Figs.~\ref{Figure3}(b) and (d). In this case the feature at zero field corresponds to the energy of the triplet states. The measured singlet triplet energy separation $E_{ST}$ is $415\mu$eV. The difference  $E_{ORB}-E_{ST}=275\mu\,$eV corresponds to the Coulomb interaction energy and is similar to what has been reported for GaAs  QDs \cite{Hanson}.

By inspecting carefully the behavior of the triplet states in Fig.~\ref{Figure3}(b), it can be seen that the triplets are not equally spaced (Fig.~\ref{Figure3}(d)); it actually seems that the three triplet states (marked by three dashed lines)  are not degenerate at $B=0T$. Ge is known to have a very strong atomistic (valence band) SOC which leads to the HH spin poiting in the perpendicular direction of  Fig.~\ref{Figure1} (c). We can envisage the triplet state as being made up of two such HH spins. This or, even more general, any triplet with an anisotropy in the  $\perp$- direction of Fig.~\ref{Figure2}(b) can be described by the following  Hamiltonian for the triplet spin ${\mathbf S}$ (see e.g.~\cite{Misiorny2013,Bruno}):
\begin{equation}
 H =-J/2 \; {\mathbf S}{\mathbf S} +g_{\perp}\mu_B{ S_\perp}{B_\perp}+ g_{\parallel}\mu_B{ S_\parallel}{B_\parallel} - DS_{\perp}^2  . \label{Eq}
\end{equation} 
Here, ${S_\perp}$ and $S_{\parallel}$ are the projections 
in the   $\perp$- and $\parallel$-direction of Fig.~\ref{Figure2}(b) and the terms of the Hamiltonian are from right to left:
The magnetic anisotropy term  $DS_{\perp}^2$ which makes it preferably  by an energy $D$ to align the triplet spin-1 in the  $\perp$-direction with strongest confinement. Its origin will be discussed in the next paragraph. The next two terms describe the Zeeman term with the magnetic field in the
two directions,  ${B_\perp}$ and ${B_\parallel}$, coupling through  different (anisotropic) $g_{\perp}$ and $g_{\parallel}$-factors. Finally, we also include the exchange term $J$ which differentiates singlet and triplet, but is not relevant in the following as we concentrate on the magnetic field dependence of the triplet states (${\mathbf S}=1$ fixed)  only. 

From the effective Hamiltonian (\ref{Eq}) for the triplet states we cannot distinguish the origin of the magnetic anisotropy. It might be due to (i) shape anisotropy caused by dipole interactions \cite{Bruno}, (ii) single ion (single quantum dot) anisotropy caused by SOC-induced transitions to excited (virtual) states
  \cite{Misiorny2013,Bruno}
or (iii) a SOC-induced anisotropic exchange $J_{\rm A}$~\cite{Burkard,Friesen}.

The last microscopic origin is certainly most natural if we think of the triplet spin-$1$ state as being made up out of two HH spin $\pm 3/2$ states, which we can  describe as two coupled pseudospin-$1/2$,  ${\mathbf S_{1} }$ and ${\mathbf S_{2} }$.  Given that these pseudospins actually describe HH  spin $\pm 3/2$ states (or the strong SOC coupling from a general perspective)  
the  coupling of these pseudospins has to be anisotropic, i.e., $H = - J{\mathbf S_{1} } {\mathbf S_{2} } -J_{\rm A}S_{1\perp}S_{2\perp}$. 
This reduces  to Eq.~(\ref{Eq})  with ${\mathbf S }={\mathbf S_{1} }+ {\mathbf S_{2} }$  and $D=J_A/2$ in the triplet subspace, up to a constant.

The eigenstates of Hamiltonian~(\ref{Eq}) can be easily calculated  and are shown in  Fig.~\ref{Figure4} for a magnetic field applied once in the  $\perp$- and once in the $\parallel$-direction. For $B=0$ the two states with $S_\perp=\pm1$ have a by $-D$ smaller energy than the third triplet state with $S_\perp=0$. Hence, the lowest triplet state is doubly degenerate and the remaining one singly degenerate in  Fig.~\ref{Figure4}. Applying now a magnetic field in the anisotropy direction  $B_\perp$ Zeeman-splits the doublet  and leaves the singly degenerate   $S_\perp=0$ state untouched (Fig.~\ref{Figure4} (a)), with $E_{S_+}-E_{S_-}=2g_\perp \mu_B B$.

 The situation with the magnetic field  $B_\parallel$ orthogonal to the anisotropy direction is somewhat more complicated. Here, for small  $B_\parallel$, the eigenstates are still predominately $S_\perp=\pm1,0$ with only a small, perturbative readmixture $\sim g_\parallel\mu B_\parallel/D$ as the magnetic field tries to align the spins in the $\parallel$-direction. This linear readmixture of the eigenstates, leads to a quadratic change of the energy eigenvalues in Fig.~\ref{Figure4} (b) for
 $g_\parallel\mu B_\parallel \ll D$.  For large   
$g_\parallel\mu B_\parallel\gg D$,  the usual Zeeman splitting of the triplet states
into   $S_\parallel =\pm 1,0$ is recovered as  the HH pseudo spins now
reorient along $B_\parallel$. This is in very good agreement with the data shown in Fig.~\ref{Figure3}(b), even though we have not adjusted the parameters but extracted these experimentally from  Fig.~\ref{Figure3}(b) and similar line traces at other $B_\parallel$'s. An even better agreement is obtained when freely adjusting $D$ and $g_\parallel$ (not shown).

These considerations clearly show that there is a ZFS and that the magnetic field dependence shows a quite different behavior for   $B_\parallel$ and   $B_\perp$.
If we have  an odd number of electrons, the doublet could also be described with Hamiltonian~(\ref{Eq}). But in this case, both $S_\perp=\pm 1/2$ states have the same anisotropy energy. Hence there is a Zeeman splitting but no ZFS as observed in  Fig.~\ref{Figure3}(a).

In order to further elucidate this behavior of the triplet state, we study in  Figures~\ref{Figure4} (c), (d) the dependence on a magnetic field  $B_{\perp}$. Using a second derivative to sharpen the features, it can be seen even more clearly that the HH triplet states are not degenerate at $B=0$. Even more, the magnetic field evolution perfectly fits with that of Fig.~\ref{Figure4}(a), which is also indicated as dashed (white) lines. Figures~\ref{Figure4} (e,f) show the same split degeneracy also for a second device. In this case orbital effects also lead to a slight bending of the states for $B_{\perp}$ and the ZFS is extracted to be $30\mu$eV. Except for this extra bending  Fig.~\ref{Figure4} (e) resembles Fig.~\ref{Figure4}(a) and  Fig.~\ref{Figure4} (f) resembles Fig.~\ref{Figure4}(b) for the two different magnetic field directions.
 From the observed splitting it is obvious that the ZFS needs to be taken into account when considering the energy band diagram of double QDs, for which it has been assumed so far, that triplet HH states are all degenerate at $B=0T$.

In conclusion, we have demonstrated the ZFS   for heavy hole states confined in a two dimensional quantum dot.  Specifically, the triplet states are split into a double and a single degenerate level.
This is not only of fundamental interest but also needs to be taken into account, for better or for worse, when operating heavy hole qubits~\cite{Watzinger2018,Hendrickx2019}.  It can be exploited for rotating and preparing a well defined quantum state using Rabi oscillations at the  ZFS (microwave) frequency, similar as for nitrogen vacancy centers in diamond \cite{Fuchs2009,Jelezko2006}. A small magnetic field can further help addressing the spin $\pm 1$ states individually. If we consider the anisotropic exchange $J_A$ as the origin of the ZFS, it can   be employed  for qubit operations \cite{Friesen} but may also be tuned (more) isotropic using proper pulse shaping  \cite{Bonesteel,Burkard}.

We acknowledge G. Burkard, C. Kloeffel, D. Loss, P. Rabl and M. Ran\v{c}i\'{c} for helpful discussions. We further acknowledge T. Adletzberger, J. Aguilera, T. Asenov, S. Bagiante, T. Menner, L. Shafeek, P. Taus, P. Traunm\"uller and D. Waldh\"ausl for their invaluable assistance. This research was supported by the Scientific Service Units of IST Austria through resources provided by the MIBA Machine Shop and the nanofabrication facility, by the FWF-P 32235 project, by the National Key R\&D Program of China (Grants No. 2016YFA0301701, 2016YFA0300600) and the European Union's Horizon 2020 research and innovation program under Grant Agreement \#862046.

\bibliography{References}

\begin{thebibliography}{36}%
\makeatletter
\providecommand \@ifxundefined [1]{%
 \@ifx{#1\undefined}
}%
\providecommand \@ifnum [1]{%
 \ifnum #1\expandafter \@firstoftwo
 \else \expandafter \@secondoftwo
 \fi
}%
\providecommand \@ifx [1]{%
 \ifx #1\expandafter \@firstoftwo
 \else \expandafter \@secondoftwo
 \fi
}%
\providecommand \natexlab [1]{#1}%
\providecommand \enquote  [1]{``#1''}%
\providecommand \bibnamefont  [1]{#1}%
\providecommand \bibfnamefont [1]{#1}%
\providecommand \citenamefont [1]{#1}%
\providecommand \href@noop [0]{\@secondoftwo}%
\providecommand \href [0]{\begingroup \@sanitize@url \@href}%
\providecommand \@href[1]{\@@startlink{#1}\@@href}%
\providecommand \@@href[1]{\endgroup#1\@@endlink}%
\providecommand \@sanitize@url [0]{\catcode `\\12\catcode `\$12\catcode
  `\&12\catcode `\#12\catcode `\^12\catcode `\_12\catcode `\%12\relax}%
\providecommand \@@startlink[1]{}%
\providecommand \@@endlink[0]{}%
\providecommand \url  [0]{\begingroup\@sanitize@url \@url }%
\providecommand \@url [1]{\endgroup\@href {#1}{\urlprefix }}%
\providecommand \urlprefix  [0]{URL }%
\providecommand \Eprint [0]{\href }%
\providecommand \doibase [0]{http://dx.doi.org/}%
\providecommand \selectlanguage [0]{\@gobble}%
\providecommand \bibinfo  [0]{\@secondoftwo}%
\providecommand \bibfield  [0]{\@secondoftwo}%
\providecommand \translation [1]{[#1]}%
\providecommand \BibitemOpen [0]{}%
\providecommand \bibitemStop [0]{}%
\providecommand \bibitemNoStop [0]{.\EOS\space}%
\providecommand \EOS [0]{\spacefactor3000\relax}%
\providecommand \BibitemShut  [1]{\csname bibitem#1\endcsname}%
\let\auto@bib@innerbib\@empty
\bibitem [{\citenamefont {Winkler}(2003)}]{Winkler}%
  \BibitemOpen
  \bibinfo {editor} {\bibfnamefont {R.}~\bibnamefont {Winkler}},\ ed.,\
  \href@noop {} {\emph {\bibinfo {title} {Spin-Orbit Coupling Effects in
  Two-Dimensional Electron and Hole Systems}}}\ (\bibinfo  {publisher}
  {Springer New York},\ \bibinfo {year} {2003})\BibitemShut {NoStop}%
\bibitem [{\citenamefont {Kloeffel}\ \emph {et~al.}(2011)\citenamefont
  {Kloeffel}, \citenamefont {Trif},\ and\ \citenamefont {Loss}}]{Kloeffel2011}%
  \BibitemOpen
  \bibfield  {author} {\bibinfo {author} {\bibfnamefont {C.}~\bibnamefont
  {Kloeffel}}, \bibinfo {author} {\bibfnamefont {M.}~\bibnamefont {Trif}}, \
  and\ \bibinfo {author} {\bibfnamefont {D.}~\bibnamefont {Loss}},\ }\href
  {\doibase https://doi.org/10.1103/PhysRevB.84.195314} {\bibfield  {journal}
  {\bibinfo  {journal} {Phys. Rev. B}\ }\textbf {\bibinfo {volume} {84}},\
  \bibinfo {pages} {195314} (\bibinfo {year} {2011})}\BibitemShut {NoStop}%
\bibitem [{\citenamefont {Hong}\ \emph {et~al.}(2018)\citenamefont {Hong},
  \citenamefont {Marcellina}, \citenamefont {Hamilton},\ and\ \citenamefont
  {Culcer}}]{Hong2018}%
  \BibitemOpen
  \bibfield  {author} {\bibinfo {author} {\bibfnamefont {L.}~\bibnamefont
  {Hong}}, \bibinfo {author} {\bibfnamefont {E.}~\bibnamefont {Marcellina}},
  \bibinfo {author} {\bibfnamefont {A.}~\bibnamefont {Hamilton}}, \ and\
  \bibinfo {author} {\bibfnamefont {D.}~\bibnamefont {Culcer}},\ }\href@noop {}
  {\bibfield  {journal} {\bibinfo  {journal} {Phys. Rev. Lett.}\ }\textbf
  {\bibinfo {volume} {121}},\ \bibinfo {pages} {087701} (\bibinfo {year}
  {2018})}\BibitemShut {NoStop}%
\bibitem [{\citenamefont {Maurand}\ \emph {et~al.}(2016)\citenamefont
  {Maurand}, \citenamefont {Jehl}, \citenamefont {Kotekar-Patil}, \citenamefont
  {Corna}, \citenamefont {Bohuslavskyi}, \citenamefont {Lavi\'eville},
  \citenamefont {Hutin}, \citenamefont {Barraud}, \citenamefont {Vinet},
  \citenamefont {Sanquer},\ and\ \citenamefont {De~Franceschi}}]{Maurand}%
  \BibitemOpen
  \bibfield  {author} {\bibinfo {author} {\bibfnamefont {R.}~\bibnamefont
  {Maurand}}, \bibinfo {author} {\bibfnamefont {X.}~\bibnamefont {Jehl}},
  \bibinfo {author} {\bibfnamefont {D.}~\bibnamefont {Kotekar-Patil}}, \bibinfo
  {author} {\bibfnamefont {A.}~\bibnamefont {Corna}}, \bibinfo {author}
  {\bibfnamefont {H.}~\bibnamefont {Bohuslavskyi}}, \bibinfo {author}
  {\bibfnamefont {R.}~\bibnamefont {Lavi\'eville}}, \bibinfo {author}
  {\bibfnamefont {L.}~\bibnamefont {Hutin}}, \bibinfo {author} {\bibfnamefont
  {S.}~\bibnamefont {Barraud}}, \bibinfo {author} {\bibfnamefont
  {M.}~\bibnamefont {Vinet}}, \bibinfo {author} {\bibfnamefont
  {M.}~\bibnamefont {Sanquer}}, \ and\ \bibinfo {author} {\bibfnamefont
  {S.}~\bibnamefont {De~Franceschi}},\ }\href@noop {} {\bibfield  {journal}
  {\bibinfo  {journal} {Nature Com.}\ }\textbf {\bibinfo {volume} {7}},\
  \bibinfo {pages} {13575} (\bibinfo {year} {2016})}\BibitemShut {NoStop}%
\bibitem [{\citenamefont {Watzinger}\ \emph {et~al.}(2018)\citenamefont
  {Watzinger}, \citenamefont {Kuku\v{c}ka}, \citenamefont {Vuku\v{s}i\'{c}},
  \citenamefont {Gao}, \citenamefont {Wang}, \citenamefont {Sch\"affler},
  \citenamefont {Zhang},\ and\ \citenamefont {Katsaros}}]{Watzinger2018}%
  \BibitemOpen
  \bibfield  {author} {\bibinfo {author} {\bibfnamefont {H.}~\bibnamefont
  {Watzinger}}, \bibinfo {author} {\bibfnamefont {J.}~\bibnamefont
  {Kuku\v{c}ka}}, \bibinfo {author} {\bibfnamefont {L.}~\bibnamefont
  {Vuku\v{s}i\'{c}}}, \bibinfo {author} {\bibfnamefont {F.}~\bibnamefont
  {Gao}}, \bibinfo {author} {\bibfnamefont {T.}~\bibnamefont {Wang}}, \bibinfo
  {author} {\bibfnamefont {F.}~\bibnamefont {Sch\"affler}}, \bibinfo {author}
  {\bibfnamefont {J.~J.}\ \bibnamefont {Zhang}}, \ and\ \bibinfo {author}
  {\bibfnamefont {G.}~\bibnamefont {Katsaros}},\ }\href@noop {} {\bibfield
  {journal} {\bibinfo  {journal} {Nature Com.}\ }\textbf {\bibinfo {volume}
  {9}},\ \bibinfo {pages} {3902} (\bibinfo {year} {2018})}\BibitemShut
  {NoStop}%
\bibitem [{\citenamefont {Hendrickx}\ \emph {et~al.}(2019)\citenamefont
  {Hendrickx}, \citenamefont {Franke}, \citenamefont {Sammak}, \citenamefont
  {Scappucci},\ and\ \citenamefont {Veldhorst}}]{Hendrickx2019}%
  \BibitemOpen
  \bibfield  {author} {\bibinfo {author} {\bibfnamefont {N.~W.}\ \bibnamefont
  {Hendrickx}}, \bibinfo {author} {\bibfnamefont {D.~P.}\ \bibnamefont
  {Franke}}, \bibinfo {author} {\bibfnamefont {A.}~\bibnamefont {Sammak}},
  \bibinfo {author} {\bibfnamefont {G.}~\bibnamefont {Scappucci}}, \ and\
  \bibinfo {author} {\bibfnamefont {M.}~\bibnamefont {Veldhorst}},\ }\href
  {\doibase arXiv:1904.11443 (2019)} {\  (\bibinfo {year} {2019}),\
  arXiv:1904.11443 (2019)}\BibitemShut {NoStop}%
\bibitem [{\citenamefont {Crippa}\ \emph {et~al.}(2019)\citenamefont {Crippa},
  \citenamefont {Ezzouch}, \citenamefont {Apr\'a}, \citenamefont {Amisse},
  \citenamefont {Lavi\'eville}, \citenamefont {Hutin}, \citenamefont
  {Bertrand}, \citenamefont {Vinet}, \citenamefont {Urdampilleta},
  \citenamefont {Meunier}, \citenamefont {Sanquer}, \citenamefont {Jehl},
  \citenamefont {Maurand},\ and\ \citenamefont {De~Franceschi}}]{Crippa2019}%
  \BibitemOpen
  \bibfield  {author} {\bibinfo {author} {\bibfnamefont {A.}~\bibnamefont
  {Crippa}}, \bibinfo {author} {\bibfnamefont {R.}~\bibnamefont {Ezzouch}},
  \bibinfo {author} {\bibfnamefont {A.}~\bibnamefont {Apr\'a}}, \bibinfo
  {author} {\bibfnamefont {A.}~\bibnamefont {Amisse}}, \bibinfo {author}
  {\bibfnamefont {R.}~\bibnamefont {Lavi\'eville}}, \bibinfo {author}
  {\bibfnamefont {L.}~\bibnamefont {Hutin}}, \bibinfo {author} {\bibfnamefont
  {B.}~\bibnamefont {Bertrand}}, \bibinfo {author} {\bibfnamefont
  {M.}~\bibnamefont {Vinet}}, \bibinfo {author} {\bibfnamefont
  {M.}~\bibnamefont {Urdampilleta}}, \bibinfo {author} {\bibfnamefont
  {T.}~\bibnamefont {Meunier}}, \bibinfo {author} {\bibfnamefont
  {M.}~\bibnamefont {Sanquer}}, \bibinfo {author} {\bibfnamefont
  {X.}~\bibnamefont {Jehl}}, \bibinfo {author} {\bibfnamefont {R.}~\bibnamefont
  {Maurand}}, \ and\ \bibinfo {author} {\bibfnamefont {S.}~\bibnamefont
  {De~Franceschi}},\ }\href@noop {} {\bibfield  {journal} {\bibinfo  {journal}
  {Nature Com.}\ }\textbf {\bibinfo {volume} {10}},\ \bibinfo {pages} {2776}
  (\bibinfo {year} {2019})}\BibitemShut {NoStop}%
\bibitem [{\citenamefont {Golovach}\ \emph {et~al.}(2006)\citenamefont
  {Golovach}, \citenamefont {Borhani},\ and\ \citenamefont {Loss}}]{Golovach}%
  \BibitemOpen
  \bibfield  {author} {\bibinfo {author} {\bibfnamefont {V.~N.}\ \bibnamefont
  {Golovach}}, \bibinfo {author} {\bibfnamefont {M.}~\bibnamefont {Borhani}}, \
  and\ \bibinfo {author} {\bibfnamefont {D.}~\bibnamefont {Loss}},\ }\href@noop
  {} {\bibfield  {journal} {\bibinfo  {journal} {Phys. Rev. B}\ }\textbf
  {\bibinfo {volume} {74}},\ \bibinfo {pages} {165319} (\bibinfo {year}
  {2006})}\BibitemShut {NoStop}%
\bibitem [{\citenamefont {Kato}\ \emph {et~al.}(2003)\citenamefont {Kato},
  \citenamefont {Myers}, \citenamefont {Driscoll}, \citenamefont {Gossard},
  \citenamefont {Levy},\ and\ \citenamefont {Awschalom}}]{Kato}%
  \BibitemOpen
  \bibfield  {author} {\bibinfo {author} {\bibfnamefont {Y.}~\bibnamefont
  {Kato}}, \bibinfo {author} {\bibfnamefont {R.~C.}\ \bibnamefont {Myers}},
  \bibinfo {author} {\bibfnamefont {D.~C.}\ \bibnamefont {Driscoll}}, \bibinfo
  {author} {\bibfnamefont {A.~C.}\ \bibnamefont {Gossard}}, \bibinfo {author}
  {\bibfnamefont {J.}~\bibnamefont {Levy}}, \ and\ \bibinfo {author}
  {\bibfnamefont {D.~D.}\ \bibnamefont {Awschalom}},\ }\href@noop {} {\bibfield
   {journal} {\bibinfo  {journal} {Science}\ }\textbf {\bibinfo {volume}
  {299}},\ \bibinfo {pages} {1201} (\bibinfo {year} {2003})}\BibitemShut
  {NoStop}%
\bibitem [{\citenamefont {Crippa}\ \emph {et~al.}(2018)\citenamefont {Crippa},
  \citenamefont {Maurand}, \citenamefont {Bourdet}, \citenamefont
  {Kotekar-Patil}, \citenamefont {Amisse}, \citenamefont {Jehl}, \citenamefont
  {Sanquer}, \citenamefont {Lavi\'eville}, \citenamefont {Bohuslavskyi},
  \citenamefont {Hutin}, \citenamefont {Barraud}, \citenamefont {Vinet},
  \citenamefont {Niquet},\ and\ \citenamefont {De~Franceschi}}]{Crippa2018}%
  \BibitemOpen
  \bibfield  {author} {\bibinfo {author} {\bibfnamefont {A.}~\bibnamefont
  {Crippa}}, \bibinfo {author} {\bibfnamefont {R.}~\bibnamefont {Maurand}},
  \bibinfo {author} {\bibfnamefont {L.}~\bibnamefont {Bourdet}}, \bibinfo
  {author} {\bibfnamefont {D.}~\bibnamefont {Kotekar-Patil}}, \bibinfo {author}
  {\bibfnamefont {A.}~\bibnamefont {Amisse}}, \bibinfo {author} {\bibfnamefont
  {X.}~\bibnamefont {Jehl}}, \bibinfo {author} {\bibfnamefont {M.}~\bibnamefont
  {Sanquer}}, \bibinfo {author} {\bibfnamefont {R.}~\bibnamefont
  {Lavi\'eville}}, \bibinfo {author} {\bibfnamefont {H.}~\bibnamefont
  {Bohuslavskyi}}, \bibinfo {author} {\bibfnamefont {L.}~\bibnamefont {Hutin}},
  \bibinfo {author} {\bibfnamefont {S.}~\bibnamefont {Barraud}}, \bibinfo
  {author} {\bibfnamefont {M.}~\bibnamefont {Vinet}}, \bibinfo {author}
  {\bibfnamefont {Y.-M.}\ \bibnamefont {Niquet}}, \ and\ \bibinfo {author}
  {\bibfnamefont {S.}~\bibnamefont {De~Franceschi}},\ }\href {\doibase
  10.1103/PhysRevLett.120.137702} {\bibfield  {journal} {\bibinfo  {journal}
  {Phys. Rev. Lett.}\ }\textbf {\bibinfo {volume} {120}},\ \bibinfo {pages}
  {137702} (\bibinfo {year} {2018})}\BibitemShut {NoStop}%
\bibitem [{\citenamefont {Vuku\v{s}i\'c}\ \emph {et~al.}(2018)\citenamefont
  {Vuku\v{s}i\'c}, \citenamefont {Kuku\v{c}ka}, \citenamefont {Watzinger},
  \citenamefont {Milem}, \citenamefont {Sch\"{a}ffler},\ and\ \citenamefont
  {Katsaros}}]{Vukusic}%
  \BibitemOpen
  \bibfield  {author} {\bibinfo {author} {\bibfnamefont {L.}~\bibnamefont
  {Vuku\v{s}i\'c}}, \bibinfo {author} {\bibfnamefont {J.}~\bibnamefont
  {Kuku\v{c}ka}}, \bibinfo {author} {\bibfnamefont {H.}~\bibnamefont
  {Watzinger}}, \bibinfo {author} {\bibfnamefont {J.~M.}\ \bibnamefont
  {Milem}}, \bibinfo {author} {\bibfnamefont {F.}~\bibnamefont
  {Sch\"{a}ffler}}, \ and\ \bibinfo {author} {\bibfnamefont {G.}~\bibnamefont
  {Katsaros}},\ }\href@noop {} {\bibfield  {journal} {\bibinfo  {journal} {Nano
  letters}\ }\textbf {\bibinfo {volume} {18}},\ \bibinfo {pages} {7141}
  (\bibinfo {year} {2018})}\BibitemShut {NoStop}%
\bibitem [{\citenamefont {Kloeffel}\ \emph {et~al.}(2018)\citenamefont
  {Kloeffel}, \citenamefont {Ran\v{c}i\'{c}},\ and\ \citenamefont
  {Loss}}]{Kloeffel2018}%
  \BibitemOpen
  \bibfield  {author} {\bibinfo {author} {\bibfnamefont {C.}~\bibnamefont
  {Kloeffel}}, \bibinfo {author} {\bibfnamefont {M.~J.}\ \bibnamefont
  {Ran\v{c}i\'{c}}}, \ and\ \bibinfo {author} {\bibfnamefont {D.}~\bibnamefont
  {Loss}},\ }\href@noop {} {\bibfield  {journal} {\bibinfo  {journal} {Phys.
  Rev. B}\ }\textbf {\bibinfo {volume} {97}},\ \bibinfo {pages} {235422}
  (\bibinfo {year} {2018})}\BibitemShut {NoStop}%
\bibitem [{\citenamefont {Luttinger}\ and\ \citenamefont
  {Kohn}(1955)}]{LuttingerKohn}%
  \BibitemOpen
  \bibfield  {author} {\bibinfo {author} {\bibfnamefont {J.~M.}\ \bibnamefont
  {Luttinger}}\ and\ \bibinfo {author} {\bibfnamefont {W.}~\bibnamefont
  {Kohn}},\ }\href@noop {} {\bibfield  {journal} {\bibinfo  {journal} {Phys.
  Rev.}\ }\textbf {\bibinfo {volume} {97}},\ \bibinfo {pages} {869} (\bibinfo
  {year} {1955})}\BibitemShut {NoStop}%
\bibitem [{\citenamefont {Luttinger}(1956)}]{Luttinger}%
  \BibitemOpen
  \bibfield  {author} {\bibinfo {author} {\bibfnamefont {J.~M.}\ \bibnamefont
  {Luttinger}},\ }\href@noop {} {\bibfield  {journal} {\bibinfo  {journal}
  {Phys. Rev.}\ }\textbf {\bibinfo {volume} {102}},\ \bibinfo {pages} {1030}
  (\bibinfo {year} {1956})}\BibitemShut {NoStop}%
\bibitem [{\citenamefont {Rau}\ \emph {et~al.}(2014)\citenamefont {Rau},
  \citenamefont {Baumann}, \citenamefont {Rusponi}, \citenamefont {Donati},
  \citenamefont {Stepanow}, \citenamefont {Gragnaniello}, \citenamefont
  {Dreiser}, \citenamefont {Piamonteze}, \citenamefont {Nolting}, \citenamefont
  {Gangopadhyay}, \citenamefont {Albertini}, \citenamefont {Macfarlane},
  \citenamefont {Lutz}, \citenamefont {Jones}, \citenamefont {Gambardella},
  \citenamefont {Heinrich},\ and\ \citenamefont {Brune}}]{Rau}%
  \BibitemOpen
  \bibfield  {author} {\bibinfo {author} {\bibfnamefont {I.~G.}\ \bibnamefont
  {Rau}}, \bibinfo {author} {\bibfnamefont {S.}~\bibnamefont {Baumann}},
  \bibinfo {author} {\bibfnamefont {S.}~\bibnamefont {Rusponi}}, \bibinfo
  {author} {\bibfnamefont {F.}~\bibnamefont {Donati}}, \bibinfo {author}
  {\bibfnamefont {S.}~\bibnamefont {Stepanow}}, \bibinfo {author}
  {\bibfnamefont {L.}~\bibnamefont {Gragnaniello}}, \bibinfo {author}
  {\bibfnamefont {J.}~\bibnamefont {Dreiser}}, \bibinfo {author} {\bibfnamefont
  {C.}~\bibnamefont {Piamonteze}}, \bibinfo {author} {\bibfnamefont
  {F.}~\bibnamefont {Nolting}}, \bibinfo {author} {\bibfnamefont
  {S.}~\bibnamefont {Gangopadhyay}}, \bibinfo {author} {\bibfnamefont {O.~R.}\
  \bibnamefont {Albertini}}, \bibinfo {author} {\bibfnamefont {R.~M.}\
  \bibnamefont {Macfarlane}}, \bibinfo {author} {\bibfnamefont {C.~P.}\
  \bibnamefont {Lutz}}, \bibinfo {author} {\bibfnamefont {B.~A.}\ \bibnamefont
  {Jones}}, \bibinfo {author} {\bibfnamefont {P.}~\bibnamefont {Gambardella}},
  \bibinfo {author} {\bibfnamefont {A.~J.}\ \bibnamefont {Heinrich}}, \ and\
  \bibinfo {author} {\bibfnamefont {H.}~\bibnamefont {Brune}},\ }\href@noop {}
  {\bibfield  {journal} {\bibinfo  {journal} {Science}\ }\textbf {\bibinfo
  {volume} {988}},\ \bibinfo {pages} {344} (\bibinfo {year}
  {2014})}\BibitemShut {NoStop}%
\bibitem [{\citenamefont {Jacobson}\ \emph {et~al.}(2015)\citenamefont
  {Jacobson}, \citenamefont {Herden}, \citenamefont {Muenks}, \citenamefont
  {Laskin}, \citenamefont {Brovko}, \citenamefont {Stepanyuk}, \citenamefont
  {Ternes},\ and\ \citenamefont {Kern}}]{Jacobson}%
  \BibitemOpen
  \bibfield  {author} {\bibinfo {author} {\bibfnamefont {P.}~\bibnamefont
  {Jacobson}}, \bibinfo {author} {\bibfnamefont {T.}~\bibnamefont {Herden}},
  \bibinfo {author} {\bibfnamefont {M.}~\bibnamefont {Muenks}}, \bibinfo
  {author} {\bibfnamefont {G.}~\bibnamefont {Laskin}}, \bibinfo {author}
  {\bibfnamefont {O.}~\bibnamefont {Brovko}}, \bibinfo {author} {\bibfnamefont
  {V.}~\bibnamefont {Stepanyuk}}, \bibinfo {author} {\bibfnamefont
  {M.}~\bibnamefont {Ternes}}, \ and\ \bibinfo {author} {\bibfnamefont
  {K.}~\bibnamefont {Kern}},\ }\href@noop {} {\bibfield  {journal} {\bibinfo
  {journal} {Nature Com.}\ }\textbf {\bibinfo {volume} {6}},\ \bibinfo {pages}
  {8536} (\bibinfo {year} {2015})}\BibitemShut {NoStop}%
\bibitem [{\citenamefont {Miyamachi}\ \emph {et~al.}(2013)\citenamefont
  {Miyamachi}, \citenamefont {Schuh}, \citenamefont {Märkl}, \citenamefont
  {Bresch}, \citenamefont {Balashov}, \citenamefont {Stöhr}, \citenamefont
  {Karlewski}, \citenamefont {Andr\'e}, \citenamefont {Marthaler},
  \citenamefont {Hoffmann}, \citenamefont {Geilhufe}, \citenamefont {Ostanin},
  \citenamefont {Hergert}, \citenamefont {Mertig}, \citenamefont {Sch\''on},
  \citenamefont {Ernst},\ and\ \citenamefont {Wulfhekel}}]{Miyamachi}%
  \BibitemOpen
  \bibfield  {author} {\bibinfo {author} {\bibfnamefont {T.}~\bibnamefont
  {Miyamachi}}, \bibinfo {author} {\bibfnamefont {T.}~\bibnamefont {Schuh}},
  \bibinfo {author} {\bibfnamefont {T.}~\bibnamefont {Märkl}}, \bibinfo
  {author} {\bibfnamefont {C.}~\bibnamefont {Bresch}}, \bibinfo {author}
  {\bibfnamefont {T.}~\bibnamefont {Balashov}}, \bibinfo {author}
  {\bibfnamefont {A.}~\bibnamefont {Stöhr}}, \bibinfo {author} {\bibfnamefont
  {C.}~\bibnamefont {Karlewski}}, \bibinfo {author} {\bibfnamefont
  {S.}~\bibnamefont {Andr\'e}}, \bibinfo {author} {\bibfnamefont
  {M.}~\bibnamefont {Marthaler}}, \bibinfo {author} {\bibfnamefont
  {M.}~\bibnamefont {Hoffmann}}, \bibinfo {author} {\bibfnamefont
  {M.}~\bibnamefont {Geilhufe}}, \bibinfo {author} {\bibfnamefont
  {S.}~\bibnamefont {Ostanin}}, \bibinfo {author} {\bibfnamefont
  {W.}~\bibnamefont {Hergert}}, \bibinfo {author} {\bibfnamefont
  {I.}~\bibnamefont {Mertig}}, \bibinfo {author} {\bibfnamefont
  {G.}~\bibnamefont {Sch\''on}}, \bibinfo {author} {\bibfnamefont
  {A.}~\bibnamefont {Ernst}}, \ and\ \bibinfo {author} {\bibfnamefont
  {W.}~\bibnamefont {Wulfhekel}},\ }\href@noop {} {\bibfield  {journal}
  {\bibinfo  {journal} {Nature}\ }\textbf {\bibinfo {volume} {503}},\ \bibinfo
  {pages} {242} (\bibinfo {year} {2013})}\BibitemShut {NoStop}%
\bibitem [{\citenamefont {Gambardella}\ \emph {et~al.}(2003)\citenamefont
  {Gambardella}, \citenamefont {Rusponi}, \citenamefont {Veronese},
  \citenamefont {Dhesi}, \citenamefont {Grazioli}, \citenamefont {Dallmeyer},
  \citenamefont {Cabria}, \citenamefont {Zeller}, \citenamefont {Dederichs},
  \citenamefont {Kern}, \citenamefont {Carbone},\ and\ \citenamefont
  {Brune}}]{Gambardella}%
  \BibitemOpen
  \bibfield  {author} {\bibinfo {author} {\bibfnamefont {P.}~\bibnamefont
  {Gambardella}}, \bibinfo {author} {\bibfnamefont {S.}~\bibnamefont
  {Rusponi}}, \bibinfo {author} {\bibfnamefont {M.}~\bibnamefont {Veronese}},
  \bibinfo {author} {\bibfnamefont {S.~S.}\ \bibnamefont {Dhesi}}, \bibinfo
  {author} {\bibfnamefont {C.}~\bibnamefont {Grazioli}}, \bibinfo {author}
  {\bibfnamefont {A.}~\bibnamefont {Dallmeyer}}, \bibinfo {author}
  {\bibfnamefont {I.}~\bibnamefont {Cabria}}, \bibinfo {author} {\bibfnamefont
  {R.}~\bibnamefont {Zeller}}, \bibinfo {author} {\bibfnamefont {P.~H.}\
  \bibnamefont {Dederichs}}, \bibinfo {author} {\bibfnamefont {K.}~\bibnamefont
  {Kern}}, \bibinfo {author} {\bibfnamefont {C.}~\bibnamefont {Carbone}}, \
  and\ \bibinfo {author} {\bibfnamefont {H.}~\bibnamefont {Brune}},\
  }\href@noop {} {\bibfield  {journal} {\bibinfo  {journal} {Science}\ }\textbf
  {\bibinfo {volume} {300}},\ \bibinfo {pages} {1130} (\bibinfo {year}
  {2003})}\BibitemShut {NoStop}%
\bibitem [{\citenamefont {Hirjibehedin}\ \emph {et~al.}(2007)\citenamefont
  {Hirjibehedin}, \citenamefont {Lin}, \citenamefont {Otte}, \citenamefont
  {Ternes}, \citenamefont {Lutz}, \citenamefont {Jones},\ and\ \citenamefont
  {Heinrich}}]{Hirjibehedin}%
  \BibitemOpen
  \bibfield  {author} {\bibinfo {author} {\bibfnamefont {C.~F.}\ \bibnamefont
  {Hirjibehedin}}, \bibinfo {author} {\bibfnamefont {C.~Y.}\ \bibnamefont
  {Lin}}, \bibinfo {author} {\bibfnamefont {A.~F.}\ \bibnamefont {Otte}},
  \bibinfo {author} {\bibfnamefont {M.}~\bibnamefont {Ternes}}, \bibinfo
  {author} {\bibfnamefont {C.~P.}\ \bibnamefont {Lutz}}, \bibinfo {author}
  {\bibfnamefont {B.~A.}\ \bibnamefont {Jones}}, \ and\ \bibinfo {author}
  {\bibfnamefont {A.~J.}\ \bibnamefont {Heinrich}},\ }\href@noop {} {\bibfield
  {journal} {\bibinfo  {journal} {Science}\ }\textbf {\bibinfo {volume}
  {317}},\ \bibinfo {pages} {1199} (\bibinfo {year} {2007})}\BibitemShut
  {NoStop}%
\bibitem [{\citenamefont {Gao}\ \emph {et~al.}(2020)\citenamefont {Gao},
  \citenamefont {Wang}, \citenamefont {Watzinger}, \citenamefont {Hu},
  \citenamefont {Rančić}, \citenamefont {Zhang}, \citenamefont {Wang},
  \citenamefont {Yao}, \citenamefont {Wang}, \citenamefont {Kukučka},
  \citenamefont {Vukušić}, \citenamefont {Kloeffel}, \citenamefont {Loss},
  \citenamefont {Liu}, \citenamefont {Katsaros},\ and\ \citenamefont
  {Zhang}}]{Gao}%
  \BibitemOpen
  \bibfield  {author} {\bibinfo {author} {\bibfnamefont {F.}~\bibnamefont
  {Gao}}, \bibinfo {author} {\bibfnamefont {J.-H.}\ \bibnamefont {Wang}},
  \bibinfo {author} {\bibfnamefont {H.}~\bibnamefont {Watzinger}}, \bibinfo
  {author} {\bibfnamefont {H.}~\bibnamefont {Hu}}, \bibinfo {author}
  {\bibfnamefont {M.~J.}\ \bibnamefont {Rančić}}, \bibinfo {author}
  {\bibfnamefont {J.-Y.}\ \bibnamefont {Zhang}}, \bibinfo {author}
  {\bibfnamefont {T.}~\bibnamefont {Wang}}, \bibinfo {author} {\bibfnamefont
  {Y.}~\bibnamefont {Yao}}, \bibinfo {author} {\bibfnamefont {G.-L.}\
  \bibnamefont {Wang}}, \bibinfo {author} {\bibfnamefont {J.}~\bibnamefont
  {Kukučka}}, \bibinfo {author} {\bibfnamefont {L.}~\bibnamefont {Vukušić}},
  \bibinfo {author} {\bibfnamefont {C.}~\bibnamefont {Kloeffel}}, \bibinfo
  {author} {\bibfnamefont {D.}~\bibnamefont {Loss}}, \bibinfo {author}
  {\bibfnamefont {F.}~\bibnamefont {Liu}}, \bibinfo {author} {\bibfnamefont
  {G.}~\bibnamefont {Katsaros}}, \ and\ \bibinfo {author} {\bibfnamefont
  {J.-J.}\ \bibnamefont {Zhang}},\ }\href {\doibase 10.1002/adma.201906523}
  {\bibfield  {journal} {\bibinfo  {journal} {Advanced Materials}\ }\textbf
  {\bibinfo {volume} {n/a}},\ \bibinfo {pages} {1906523} (\bibinfo {year}
  {2020})}\BibitemShut {NoStop}%
\bibitem [{\citenamefont {Zhang}\ \emph {et~al.}(2012)\citenamefont {Zhang},
  \citenamefont {Katsaros}, \citenamefont {Montalenti}, \citenamefont
  {Scopece}, \citenamefont {Rezaev}, \citenamefont {Mickel}, \citenamefont
  {Rellinghaus}, \citenamefont {Miglio}, \citenamefont {De~Franceschi},
  \citenamefont {Rastelli},\ and\ \citenamefont {Schmidt}}]{Zhang}%
  \BibitemOpen
  \bibfield  {author} {\bibinfo {author} {\bibfnamefont {J.~J.}\ \bibnamefont
  {Zhang}}, \bibinfo {author} {\bibfnamefont {G.}~\bibnamefont {Katsaros}},
  \bibinfo {author} {\bibfnamefont {F.}~\bibnamefont {Montalenti}}, \bibinfo
  {author} {\bibfnamefont {D.}~\bibnamefont {Scopece}}, \bibinfo {author}
  {\bibfnamefont {R.~O.}\ \bibnamefont {Rezaev}}, \bibinfo {author}
  {\bibfnamefont {C.}~\bibnamefont {Mickel}}, \bibinfo {author} {\bibfnamefont
  {B.}~\bibnamefont {Rellinghaus}}, \bibinfo {author} {\bibfnamefont
  {L.}~\bibnamefont {Miglio}}, \bibinfo {author} {\bibfnamefont
  {S.}~\bibnamefont {De~Franceschi}}, \bibinfo {author} {\bibfnamefont
  {A.}~\bibnamefont {Rastelli}}, \ and\ \bibinfo {author} {\bibfnamefont
  {O.~G.}\ \bibnamefont {Schmidt}},\ }\href {\doibase
  10.1103/PhysRevLett.109.085502} {\bibfield  {journal} {\bibinfo  {journal}
  {Phys. Rev. Lett.}\ }\textbf {\bibinfo {volume} {109}},\ \bibinfo {pages}
  {085502} (\bibinfo {year} {2012})}\BibitemShut {NoStop}%
\bibitem [{\citenamefont {Katsaros}\ \emph {et~al.}(2011)\citenamefont
  {Katsaros}, \citenamefont {Golovach}, \citenamefont {Spathis}, \citenamefont
  {Ares}, \citenamefont {Stoffel}, \citenamefont {Fournel}, \citenamefont
  {Schmidt}, \citenamefont {Glazman},\ and\ \citenamefont
  {De~Franceschi}}]{Katsaros2011}%
  \BibitemOpen
  \bibfield  {author} {\bibinfo {author} {\bibfnamefont {G.}~\bibnamefont
  {Katsaros}}, \bibinfo {author} {\bibfnamefont {V.~N.}\ \bibnamefont
  {Golovach}}, \bibinfo {author} {\bibfnamefont {P.}~\bibnamefont {Spathis}},
  \bibinfo {author} {\bibfnamefont {N.}~\bibnamefont {Ares}}, \bibinfo {author}
  {\bibfnamefont {M.}~\bibnamefont {Stoffel}}, \bibinfo {author} {\bibfnamefont
  {F.}~\bibnamefont {Fournel}}, \bibinfo {author} {\bibfnamefont {O.~G.}\
  \bibnamefont {Schmidt}}, \bibinfo {author} {\bibfnamefont {L.~I.}\
  \bibnamefont {Glazman}}, \ and\ \bibinfo {author} {\bibfnamefont
  {S.}~\bibnamefont {De~Franceschi}},\ }\href {\doibase
  doi.org/10.1103/PhysRevLett.107.246601} {\bibfield  {journal} {\bibinfo
  {journal} {Phys. Rev. Lett.}\ }\textbf {\bibinfo {volume} {107}},\ \bibinfo
  {pages} {246601} (\bibinfo {year} {2011})}\BibitemShut {NoStop}%
\bibitem [{\citenamefont {Watzinger}\ \emph {et~al.}(2016)\citenamefont
  {Watzinger}, \citenamefont {Kloeffel}, \citenamefont {Vuku\v{s}i\'c},
  \citenamefont {Rossell}, \citenamefont {Sessi}, \citenamefont {Kuku\v{c}ka},
  \citenamefont {Kirchschlager}, \citenamefont {Lausecker}, \citenamefont
  {Truhlar}, \citenamefont {Glaser}, \citenamefont {Rastelli}, \citenamefont
  {Fuhrer}, \citenamefont {Loss},\ and\ \citenamefont
  {Katsaros}}]{Watzinger2016}%
  \BibitemOpen
  \bibfield  {author} {\bibinfo {author} {\bibfnamefont {H.}~\bibnamefont
  {Watzinger}}, \bibinfo {author} {\bibfnamefont {C.}~\bibnamefont {Kloeffel}},
  \bibinfo {author} {\bibfnamefont {L.}~\bibnamefont {Vuku\v{s}i\'c}}, \bibinfo
  {author} {\bibfnamefont {M.~D.}\ \bibnamefont {Rossell}}, \bibinfo {author}
  {\bibfnamefont {V.}~\bibnamefont {Sessi}}, \bibinfo {author} {\bibfnamefont
  {J.}~\bibnamefont {Kuku\v{c}ka}}, \bibinfo {author} {\bibfnamefont
  {R.}~\bibnamefont {Kirchschlager}}, \bibinfo {author} {\bibfnamefont
  {E.}~\bibnamefont {Lausecker}}, \bibinfo {author} {\bibfnamefont
  {A.}~\bibnamefont {Truhlar}}, \bibinfo {author} {\bibfnamefont
  {M.}~\bibnamefont {Glaser}}, \bibinfo {author} {\bibfnamefont
  {A.}~\bibnamefont {Rastelli}}, \bibinfo {author} {\bibfnamefont
  {A.}~\bibnamefont {Fuhrer}}, \bibinfo {author} {\bibfnamefont
  {D.}~\bibnamefont {Loss}}, \ and\ \bibinfo {author} {\bibfnamefont
  {G.}~\bibnamefont {Katsaros}},\ }\href@noop {} {\bibfield  {journal}
  {\bibinfo  {journal} {Nano letters}\ }\textbf {\bibinfo {volume} {16}},\
  \bibinfo {pages} {6879} (\bibinfo {year} {2016})}\BibitemShut {NoStop}%
\bibitem [{\citenamefont {Kogan}\ \emph {et~al.}(2004)\citenamefont {Kogan},
  \citenamefont {Amasha}, \citenamefont {Goldhaber-Gordon}, \citenamefont
  {Granger}, \citenamefont {Kastner},\ and\ \citenamefont {Shtrikman}}]{Kogan}%
  \BibitemOpen
  \bibfield  {author} {\bibinfo {author} {\bibfnamefont {A.}~\bibnamefont
  {Kogan}}, \bibinfo {author} {\bibfnamefont {S.}~\bibnamefont {Amasha}},
  \bibinfo {author} {\bibfnamefont {D.}~\bibnamefont {Goldhaber-Gordon}},
  \bibinfo {author} {\bibfnamefont {G.}~\bibnamefont {Granger}}, \bibinfo
  {author} {\bibfnamefont {M.~A.}\ \bibnamefont {Kastner}}, \ and\ \bibinfo
  {author} {\bibfnamefont {H.}~\bibnamefont {Shtrikman}},\ }\href {\doibase
  10.1103/PhysRevLett.93.166602} {\bibfield  {journal} {\bibinfo  {journal}
  {Phys. Rev. Lett.}\ }\textbf {\bibinfo {volume} {93}},\ \bibinfo {pages}
  {166602} (\bibinfo {year} {2004})}\BibitemShut {NoStop}%
\bibitem [{\citenamefont {De~Franceschi}\ \emph {et~al.}(2001)\citenamefont
  {De~Franceschi}, \citenamefont {Sasaki}, \citenamefont {Elzerman},
  \citenamefont {van~der Wiel}, \citenamefont {Tarucha},\ and\ \citenamefont
  {Kouwenhoven}}]{Franceschi}%
  \BibitemOpen
  \bibfield  {author} {\bibinfo {author} {\bibfnamefont {S.}~\bibnamefont
  {De~Franceschi}}, \bibinfo {author} {\bibfnamefont {S.}~\bibnamefont
  {Sasaki}}, \bibinfo {author} {\bibfnamefont {J.~M.}\ \bibnamefont
  {Elzerman}}, \bibinfo {author} {\bibfnamefont {W.~G.}\ \bibnamefont {van~der
  Wiel}}, \bibinfo {author} {\bibfnamefont {S.}~\bibnamefont {Tarucha}}, \ and\
  \bibinfo {author} {\bibfnamefont {L.~P.}\ \bibnamefont {Kouwenhoven}},\
  }\href {\doibase 10.1103/PhysRevLett.86.878} {\bibfield  {journal} {\bibinfo
  {journal} {Phys. Rev. Lett.}\ }\textbf {\bibinfo {volume} {86}},\ \bibinfo
  {pages} {878} (\bibinfo {year} {2001})}\BibitemShut {NoStop}%
\bibitem [{\citenamefont {Katsaros}\ \emph {et~al.}(2010)\citenamefont
  {Katsaros}, \citenamefont {Spathis}, \citenamefont {Stoffel}, \citenamefont
  {Fournel}, \citenamefont {Mongillo}, \citenamefont {Bouchiat}, \citenamefont
  {Lefloch}, \citenamefont {Rastelli}, \citenamefont {Schmidt},\ and\
  \citenamefont {De~Franceschi}}]{Katsaros2010}%
  \BibitemOpen
  \bibfield  {author} {\bibinfo {author} {\bibfnamefont {G.}~\bibnamefont
  {Katsaros}}, \bibinfo {author} {\bibfnamefont {P.}~\bibnamefont {Spathis}},
  \bibinfo {author} {\bibfnamefont {M.}~\bibnamefont {Stoffel}}, \bibinfo
  {author} {\bibfnamefont {F.}~\bibnamefont {Fournel}}, \bibinfo {author}
  {\bibfnamefont {M.}~\bibnamefont {Mongillo}}, \bibinfo {author}
  {\bibfnamefont {V.}~\bibnamefont {Bouchiat}}, \bibinfo {author}
  {\bibfnamefont {F.}~\bibnamefont {Lefloch}}, \bibinfo {author} {\bibfnamefont
  {A.}~\bibnamefont {Rastelli}}, \bibinfo {author} {\bibfnamefont {O.~G.}\
  \bibnamefont {Schmidt}}, \ and\ \bibinfo {author} {\bibfnamefont
  {S.}~\bibnamefont {De~Franceschi}},\ }\href {\doibase
  doi:10.1038/nnano.2010.84} {\bibfield  {journal} {\bibinfo  {journal} {Nature
  Nanotechnology}\ }\textbf {\bibinfo {volume} {5}},\ \bibinfo {pages} {458}
  (\bibinfo {year} {2010})}\BibitemShut {NoStop}%
\bibitem [{\citenamefont {Nenashev}\ \emph {et~al.}(2003)\citenamefont
  {Nenashev}, \citenamefont {Dvurechenskii},\ and\ \citenamefont
  {Zinovieva}}]{Nenashev}%
  \BibitemOpen
  \bibfield  {author} {\bibinfo {author} {\bibfnamefont {A.}~\bibnamefont
  {Nenashev}}, \bibinfo {author} {\bibfnamefont {A.}~\bibnamefont
  {Dvurechenskii}}, \ and\ \bibinfo {author} {\bibfnamefont {A.}~\bibnamefont
  {Zinovieva}},\ }\href@noop {} {\bibfield  {journal} {\bibinfo  {journal}
  {Phys. Rev. B}\ }\textbf {\bibinfo {volume} {67}},\ \bibinfo {pages} {205301}
  (\bibinfo {year} {2003})}\BibitemShut {NoStop}%
\bibitem [{\citenamefont {Hanson}\ \emph {et~al.}(2007)\citenamefont {Hanson},
  \citenamefont {Kouwenhoven}, \citenamefont {Petta}, \citenamefont {Tarucha},\
  and\ \citenamefont {Vandersypen}}]{Hanson}%
  \BibitemOpen
  \bibfield  {author} {\bibinfo {author} {\bibfnamefont {R.}~\bibnamefont
  {Hanson}}, \bibinfo {author} {\bibfnamefont {L.~P.}\ \bibnamefont
  {Kouwenhoven}}, \bibinfo {author} {\bibfnamefont {J.~R.}\ \bibnamefont
  {Petta}}, \bibinfo {author} {\bibfnamefont {S.}~\bibnamefont {Tarucha}}, \
  and\ \bibinfo {author} {\bibfnamefont {L.~M.~K.}\ \bibnamefont
  {Vandersypen}},\ }\href {\doibase 10.1103/RevModPhys.79.1217} {\bibfield
  {journal} {\bibinfo  {journal} {Rev. Mod. Phys.}\ }\textbf {\bibinfo {volume}
  {79}},\ \bibinfo {pages} {1217} (\bibinfo {year} {2007})}\BibitemShut
  {NoStop}%
\bibitem [{\citenamefont {Kouwenhoven}\ \emph {et~al.}(2001)\citenamefont
  {Kouwenhoven}, \citenamefont {Austing},\ and\ \citenamefont
  {Tarucha}}]{Kouwenhoven}%
  \BibitemOpen
  \bibfield  {author} {\bibinfo {author} {\bibfnamefont {L.~P.}\ \bibnamefont
  {Kouwenhoven}}, \bibinfo {author} {\bibfnamefont {D.~G.}\ \bibnamefont
  {Austing}}, \ and\ \bibinfo {author} {\bibfnamefont {S.}~\bibnamefont
  {Tarucha}},\ }\href@noop {} {\bibfield  {journal} {\bibinfo  {journal} {Rep.
  Prog. Phys.}\ }\textbf {\bibinfo {volume} {64}},\ \bibinfo {pages} {701}
  (\bibinfo {year} {2001})}\BibitemShut {NoStop}%
\bibitem [{Mis()}]{Misiorny2013}%
  \BibitemOpen
  \href@noop {} {\bibinfo  {journal} {See e.g.\ M. Misiorny, M. Hell, and M. R.
  Wegewijs Nature Physics {\bf 9}, 801 (2013)}\ }\BibitemShut {NoStop}%
\bibitem [{Bru()}]{Bruno}%
  \BibitemOpen
\bibfield  {journal} {  }\href@noop {} {\bibinfo  {journal} {For a pedagogical
  discussion of various magnetic anisotropy terms, see P. Bruno, Physical
  Origins and Theoretical Models of Magnetic Anisotropy, in 24. IFF-Ferienkurs
  Forschungszentrum J\"ulich (Forschungszentrum J\"ulich, J\"ulich, 1993); ISBN
  3893361103}\ }\BibitemShut {NoStop}%
\bibitem [{\citenamefont {Burkard}\ and\ \citenamefont {Loss}(2002)}]{Burkard}%
  \BibitemOpen
\bibfield  {journal} {  }\bibfield  {author} {\bibinfo {author} {\bibfnamefont
  {G.}~\bibnamefont {Burkard}}\ and\ \bibinfo {author} {\bibfnamefont
  {D.}~\bibnamefont {Loss}},\ }\href {\doibase 10.1103/PhysRevLett.88.047903}
  {\bibfield  {journal} {\bibinfo  {journal} {Phys. Rev. Lett.}\ }\textbf
  {\bibinfo {volume} {88}},\ \bibinfo {pages} {047903} (\bibinfo {year}
  {2002})}\BibitemShut {NoStop}%
\bibitem [{\citenamefont {Shim}\ \emph {et~al.}(2011)\citenamefont {Shim},
  \citenamefont {Oh}, \citenamefont {Hu},\ and\ \citenamefont
  {Friesen}}]{Friesen}%
  \BibitemOpen
  \bibfield  {author} {\bibinfo {author} {\bibfnamefont {Y.-P.}\ \bibnamefont
  {Shim}}, \bibinfo {author} {\bibfnamefont {S.}~\bibnamefont {Oh}}, \bibinfo
  {author} {\bibfnamefont {X.}~\bibnamefont {Hu}}, \ and\ \bibinfo {author}
  {\bibfnamefont {M.}~\bibnamefont {Friesen}},\ }\href@noop {} {\bibfield
  {journal} {\bibinfo  {journal} {Phys. Rev. Lett.}\ }\textbf {\bibinfo
  {volume} {106}},\ \bibinfo {pages} {180503} (\bibinfo {year}
  {2011})}\BibitemShut {NoStop}%
\bibitem [{\citenamefont {Fuchs}\ \emph {et~al.}(2009)\citenamefont {Fuchs},
  \citenamefont {Dobrovitski}, \citenamefont {Toyli}, \citenamefont
  {Heremans},\ and\ \citenamefont {Awschalom}}]{Fuchs2009}%
  \BibitemOpen
  \bibfield  {author} {\bibinfo {author} {\bibfnamefont {G.~D.}\ \bibnamefont
  {Fuchs}}, \bibinfo {author} {\bibfnamefont {V.~V.}\ \bibnamefont
  {Dobrovitski}}, \bibinfo {author} {\bibfnamefont {D.~M.}\ \bibnamefont
  {Toyli}}, \bibinfo {author} {\bibfnamefont {F.~J.}\ \bibnamefont {Heremans}},
  \ and\ \bibinfo {author} {\bibfnamefont {D.~D.}\ \bibnamefont {Awschalom}},\
  }\href {\doibase 10.1126/science.1181193} {\bibfield  {journal} {\bibinfo
  {journal} {Science}\ }\textbf {\bibinfo {volume} {326}},\ \bibinfo {pages}
  {1520} (\bibinfo {year} {2009})},\ \Eprint
  {http://arxiv.org/abs/https://science.sciencemag.org/content/326/5959/1520.full.pdf}
  {https://science.sciencemag.org/content/326/5959/1520.full.pdf} \BibitemShut
  {NoStop}%
\bibitem [{\citenamefont {Jelezko}\ and\ \citenamefont
  {Wrachtrup}(2006)}]{Jelezko2006}%
  \BibitemOpen
  \bibfield  {author} {\bibinfo {author} {\bibfnamefont {F.}~\bibnamefont
  {Jelezko}}\ and\ \bibinfo {author} {\bibfnamefont {J.}~\bibnamefont
  {Wrachtrup}},\ }\href {\doibase 10.1002/pssa.200671403} {\bibfield  {journal}
  {\bibinfo  {journal} {physica status solidi (a)}\ }\textbf {\bibinfo {volume}
  {203}},\ \bibinfo {pages} {3207} (\bibinfo {year} {2006})}\BibitemShut
  {NoStop}%
\bibitem [{\citenamefont {Bonesteel}\ \emph {et~al.}(2001)\citenamefont
  {Bonesteel}, \citenamefont {Stepanenko},\ and\ \citenamefont
  {DiVincenzo}}]{Bonesteel}%
  \BibitemOpen
  \bibfield  {author} {\bibinfo {author} {\bibfnamefont {N.~E.}\ \bibnamefont
  {Bonesteel}}, \bibinfo {author} {\bibfnamefont {D.}~\bibnamefont
  {Stepanenko}}, \ and\ \bibinfo {author} {\bibfnamefont {D.~P.}\ \bibnamefont
  {DiVincenzo}},\ }\href {\doibase 10.1103/PhysRevLett.87.207901} {\bibfield
  {journal} {\bibinfo  {journal} {Phys. Rev. Lett.}\ }\textbf {\bibinfo
  {volume} {87}},\ \bibinfo {pages} {207901} (\bibinfo {year}
  {2001})}\BibitemShut {NoStop}%
\end{thebibliography}%

\end{document}